\documentclass[prb,twocolumn,showpacs]{revtex4}     
\usepackage{epsfig,amsmath} 

\begin{document}

\title{Nernst effect of Dirac fermions in graphene under weak magnetic field} 

\author{Xin-Zhong Yan$^{1}$ and C. S. Ting$^2$}
\affiliation{$^{1}$Institute of Physics, Chinese Academy of Sciences, P.O. Box 603, 
Beijing 100190, China\\
$^{2}$Texas Center for Superconductivity, University of Houston, Houston, Texas 77204, USA}
 
\date{\today}
 
\begin{abstract}
The derivation for the transport coefficients of an electron system in the presence of temperature gradient and the electric and magnetic fields are presented. The Nernst conductivity and the transverse thermoelectric power of the Dirac fermions in graphene under charged impurity scatterings and weak magnetic field are calculated on basis of the self-consistent Born approximation. The result is compared with so far the available experimental data.  
\end{abstract}


\maketitle

\section{Introduction}

The Nernst effect as well as the thermoelectric power is a sensitive probe of the impurity scatterings in an electron system. Recently, the Hall and the Nernst effects in graphene have been studied experimentally at relatively strong \cite{Zuev,Wei,Ong} and moderate \cite{Wei} magnetic fields. Graphene in most of the experiment devices is absorbed on the surface of SiO$_2$. There are strong evidences that the charged impurities in the substrate are responsible for the carrier density dependences of the electric conductivity \cite{Ando1,Cheianov,Tan,Nomura,Hwang,Yan,Chen} and the Hall coefficient \cite{Yan1} as measured in the experiments by Novoselov {\it et al.}.\cite{Geim} At strong magnetic field, the carriers are in the Landau quantized states. In the interior of the system under the strong magnetic field, the carriers are mostly localized around the charged impurities. Though, the Hall effect seems weakly dependent of the impurity scatterings because the current is most likely conducted by the edge states that are not localized.\cite{Goerbig} The standard Green's function theory of many body system has difficulty to treat the charge transport under scatterings of the charged impurities in a strong magnetic field since it deals with the bulk states of the electrons. On the other hand, at weak magnetic field when the effect of Landau quantization is negligible, the standard Green's function theory should be applicable for investigating the magnetothermoelectric transports of the electron system. 

Based on the self-consistent Born approximation (SCBA) for Dirac fermions under the charged impurity scatterings, we have recently developed an electronic transport formalism for graphene.\cite{Yan,Yan1,Yan2} It has been shown that the experimentally measured electric conductivity, the inverse Hall coefficient\cite{Geim} and the thermoelectric power\cite{Zuev,Wei,Ong} are successfully explained by our approach. In this work, along the same approach, we study the Nernst effect of Dirac fermions as a function of the carrier density under a weak magnetic field. Though there exists no experimental measurements of the Nernst effect in a weak field so far, we show that our obtained results  for the transverse thermoelectric power could qualitatively compare with the experimental measurements \cite{Wei} in magnetic fields of moderate strength. We intend to examine to what extend the theory is valid in dealing with the transport properties of graphene.

Meanwhile in doing this work, we present a derivation of the transport coefficients of an electron system under the temperature gradient and the electric and magnetic fields being applied. 

The model of electrons in graphene is established from its energy band structure in the first Brillouin zone corresponding to a honeycomb lattice. At low carrier concentration, the low energy excitations of electrons in graphene can be viewed as massless Dirac fermions.\cite{Wallace,Yan4,Ando,Castro,McCann,Geim,Zhang} That is, the energy linearly depends on the momentum around the two Dirac points in the first Brillouin zone. Using the Pauli matrices $\sigma$'s and $\tau$'s to coordinate the electrons in the two sublattices ($a$ and $b$) of the honeycomb lattice and two valleys (around the two Dirac points 1 and 2), respectively, and suppressing the spin indices for briefness, the Hamiltonian of the system is given by
\begin{equation}
H = \sum_{k}\psi^{\dagger}_{k}[v\vec
 k\cdot\vec\sigma\tau_z-\mu]\psi_{k}+\frac{1}{V}\sum_{kq}V_i(q)\psi^{\dagger}_{k-q}\psi_{k} \label{H}
\end{equation}
where $\psi^{\dagger}_{k}=(c^{\dagger}_{ka1},c^{\dagger}_{kb1},c^{\dagger}_{kb2},c^{\dagger}_{ka2})$ is the fermion operator, the momentum $k$ is measured from the center of each valley, $v$ ($\sim$ 5.86 eV\AA) is the velocity of electrons, $\mu$ is the chemical potential, $V$ is the volume of system, and $V_i(q) = n_i(-q)v_0(q)$ is the charged impurity potential.\cite{Yan1,Yan2} Here, $n_i(-q)$ is the impurity density and $v_0(q)$ is given by the Thomas-Fermi (TF) type
\begin{equation}
v_0(q)=2\pi e^2/(q+q_{TF})\epsilon
\end{equation}
where $q_{TF} = 4k_Fe^2/v\epsilon$ is the TF wavenumber, $k_F = \sqrt{\pi n}$ (with $n$ as the carrier density) is the Fermi wavenumber, and $\epsilon \sim 3$ is the effective dielectric constant. For briefness, we hereafter use units of $v = \hbar = k_B ({\rm the~ Boltzmann~ constant})= 1$.

With SCBA,\cite{Fradkin,Lee1} the Green's function 
\begin{eqnarray}
G(k,\omega)&=&[\omega+\mu-\vec
 k\cdot\vec\sigma\tau_z-\Sigma(k,\omega)]^{-1}\nonumber\\
&\equiv& g_0(k,\omega)+ g_c(k,\omega)\hat k\cdot\vec\sigma\tau_z\nonumber 
\end{eqnarray}
and the self-energy $\Sigma(k,\omega)$ of the single particles are determined by coupled integral equations.\cite{Yan} The diagonal and off diagonal parts, $g_0$ and $g_c$ respectively, of the Green's function can be expressed as 
\begin{eqnarray}
g_{0,c}(k,\omega) = [g_+(k,\omega)\pm g_-(k,\omega)]/2 \nonumber
\end{eqnarray}
with $g_{\pm}$ as the upper and lower band Green's functions. Corresponding to the SCBA to the self-energy, the current vertex correction $\Gamma_x(k,\omega_1,\omega_2)$ is given by the ladder-diagrams approximation as shown in Fig. 1(a). $\Gamma_x(k,\omega_1,\omega_2)$ is expanded as
\begin{equation}
\Gamma_x(k,\omega_1,\omega_2)=\sum_{j=0}^3y_j(k,\omega_1,\omega_2)A^x_j(\hat k) \label{vt}
\end{equation}
where $A^x_0(\hat k)=\tau_z\sigma_x$, $A^x_1(\hat k)=\sigma_x\vec\sigma\cdot\hat k$, $A^x_2(\hat k)=\vec\sigma\cdot\hat k\sigma_x$, $A^x_3(\hat k)=\tau_z\vec\sigma\cdot\hat k\sigma_x\vec\sigma\cdot\hat k$, and $y_j(k,\omega_1,\omega_2)$ are determined by four-coupled integral equations. \cite{Yan} The functions $y_j$ describe how the current vertex is renormalized by the impurity scatterings from the bare one $A^x_0(\hat k)$. 

\section{Formalism}

\subsection{General formula of the transport coefficients}

To study the Nernst effect of graphene, we consider the electronic transport of Dirac fermions under weak in-plane temperature gradient $\nabla T$, electric potential $\phi$, and weak magnetic filed $\vec B = \nabla\times\vec A$ perpendicular to the graphene plane. Here, $\vec A$ is the vector potential. Since the temperature gradient $\nabla T$ is not a dynamic quantity, we cannot directly apply the linear response theory (LRT) to treat the current response to $\nabla T$. To use LRT, one usually introduces a fictitious gravitational potential (that couples with the Hamiltonian) following the work of Luttinger and obtains the transport coefficients using the Einstein argument relating the currents response to the external perturbations.\cite{Luttinger} Here, we present a derivation from a microscopic point of view.

First, following the idea of Luttinger,\cite{Luttinger} suppose the system with a variable temperature $T(r)$ in locally equilibrium everywhere in space. Specifically, consider the system is divided into small cells but microscopically large enough. The Hamiltonian of the cell at $r_j$ with chemical potential $\mu(r_j)$ and temperature $T(r_j)$ are given by $H_j = \int_{{\rm cell}-j} d\vec r \psi^{\dagger}(r)[h(r)-\mu(r_j)]\psi(r)$ with $h(r) = \vec\sigma\tau_z\cdot(-i\nabla+\vec A)+V_i(r)$ and $\psi(r)$ as the operator of Dirac fermions in real space. The distribution function of the system is then given by 
\begin{eqnarray}
\rho_1 &=& Z^{-1}\exp[-\sum_{j}H_j/T(r_j)] \label{dstr}
\end{eqnarray}
where $Z$ is the normalization constant. Instead of considering $T(r)$, we intend to find out an equivalent system determined by an effective Hamiltonian $H_{eff}$ at constant temperature $T_0$. Its distribution function is
\begin{eqnarray}
\rho_2 &=& Z^{-1}\exp(-H_{eff}/T_0). \label{tdstr}
\end{eqnarray}
From $\rho_1 = \rho_2$, we have $H_{eff} = \sum_{j}H_jT_0/T(r_j)$. Suppose each cell is macroscopically so small enough that the summation can be replaced with integral over space. $H_{eff}$ reads 
\begin{eqnarray}
H_{eff} &=& \int d\vec r \psi^{\dagger}(r)[h(r)-\mu(r)]\circ\frac{T_0}{T(r)}\psi(r)  \nonumber\\
&=& \int d\vec r \psi^{\dagger}(r)[h(r)+h(r)\circ\Psi(r)+\Phi(r)-\mu_0]\psi(r)  \nonumber\\
&\equiv& H[\Phi,\Psi] \nonumber
\end{eqnarray}
with $h(r)\circ\Psi(r) = \{h(r),\Psi(r)\}/2$ (here $\{A,B\}$ is the anti-commutation relation between $A$ and $B$) and
\begin{eqnarray}
\Phi(r)&=& \mu_0-T_0\mu(r)/T(r),  \label{phi0}\\
\Psi(r)&=& T_0/T(r)-1. \label{psi0}
\end{eqnarray}
Here, $T_0$ and $\mu_0$ are the average temperature and chemical potential, respectively. In the limit $T(r) \to T_0$, we have $[\mu(r),\Phi(r),\Psi(r)] \to [\mu_0,0,0]$. By so doing, the system with variable temperature $T(r)$ in the local equilibrium state is now described by an equivalent one under the potentials $[\Phi(r),\Psi(r)]$ at constant $T_0$ and $\mu_0$. 

Now we go back to the original problem: How do the currents respond to the temperature gradient $\nabla T$? Initially the system is in the equilibrium state of $H[0,0]$. With gradually turning on $\nabla T$, the system $H[0,0]$ becomes unstable because the equilibrium state shifts to $H[\Phi(r),\Psi(r)]$ and thereby currents are produced. \{In the shifting process $[0,0] \to [\Phi(r),\Psi(r)]$, $T_0$ and $\mu_0$ are kept as constants.\} The perturbation here is the difference $H[0,0]-H[\Phi(r),\Psi(r)]$. This is different from the usual case that the perturbations are due to applying the external dynamic potentials and the initial equilibrium state given by $H[0,0]$ keeps unchanged. 

Generally, in addition to $\nabla T$, with the external scalar potential $\phi$ being applied, the system under consideration is $H[\phi,0]$. With respect to the equilibrium system $H[\Phi,\Psi]$, the perturbation is $H[\phi,0]-H[\Phi(r),\Psi(r)]$. Mathematically, we have
\begin{eqnarray}
H[\phi,0] &=& H[\Phi,\Psi]+ H[\phi,0]-H[\Phi,\Psi] \nonumber\\
&\equiv& H[\Phi,\Psi]+ H' \nonumber
\end{eqnarray}
with $H'$ given by
\begin{eqnarray}
H' &=& \int d\vec r \psi^{\dagger}(r)[-h(r)\circ\Psi(r)+\phi(r)-\Phi(r)]\psi(r)  \nonumber\\
&\equiv& \int d\vec r \psi^{\dagger}(r)[\Phi_1(r)+\xi(r)\circ\Phi_2(r)]\psi(r)  \nonumber
\end{eqnarray}
where $\Phi_1(r) = \phi(r)-\Phi(r)-\mu_0\Psi(r)$, $\Phi_2(r) = -\Psi(r)$, and $\xi(r)=h(r)-\mu_0$. Here $\Phi_1(r)$ and $\Phi_2(r)$ take the role as the perturbation potentials. In the limit $\nabla T \to 0$, the negtive forces $\nabla\Phi_1(r)$ and $\nabla\Psi_1(r)$ read
\begin{eqnarray}
\nabla\Phi_1(r)&=& \nabla[\phi(r)-\mu(r)]= e\vec E ,  \label{phi1}\\
\nabla\Phi_2(r)&=& -T_0\nabla[1/T(r)]. \label{psi1}
\end{eqnarray}
Hereafter we denote $\mu_0$ and $T_0$ simply as $\mu$ and $T$, respectively for briefness. According to LRT, we need to find out the corresponding currents determined from the equations of continuity. We here consider the relevant currents.

(i) For the potentials $\Phi_1$ and $\Phi_2$ (actually the corresponding vector potentials), the coupling currents to be determined are $J_1(r)$ and $J_2(r)$, respectively. One may consider to convert the facts $\psi^{\dagger}(r)\psi(r)$ (coupled to $\Phi_1$ in $H'$) and $\psi^{\dagger}(r)\xi(r)\psi(r)$ (coupled to $\Phi_2$) into the respective currents in the picture of $H[\Phi,\Psi]$ (from which the perturbed system evolve). The resulted currents then contain the terms of $\Phi_1$ and $\Phi_2$. However, since $H'$ is already linear in $\Phi_1$ and $\Phi_2$, we just only need to consider them in the picture of $H[0,0]$. For the unperturbed system $H[0,0]$, they are the particle current and the heat current,
\begin{eqnarray}
\vec J_1(r)&\equiv& \vec J(r)= \psi^{\dagger}(r)\vec j\psi(r),  \label{crt1} \\
\vec J_2(r)&\equiv& \vec J^Q(r)= \psi^{\dagger}(r)\xi(r)\circ\vec j\psi(r).  \label{crt2} 
\end{eqnarray}
with $\vec j = \vec\sigma\tau_z$. 

(ii) The currents of the reference (using subscript r) system $H[\Phi,\Psi]$ itself can be obtained from the known results of the particle current $\vec J_{1r}$ and energy current $\vec J^E_r$ in Ref. \onlinecite{Luttinger}. The results are
\begin{eqnarray}
\vec J_{1r}(r)&=& \vec J(r)\circ[1+\Psi(r)], \nonumber \\
\vec J_{2r}(r)&=& \vec J^E_r(r)-\mu\vec J_{1r}(r)\nonumber\\
&=& \vec J^Q(r)\circ[1+2\Psi(r)]+[\Phi(r)+\mu\Psi(r)]\circ\vec J(r).  \nonumber
\end{eqnarray}
Their averages under $H[\Phi,\Psi]$ vanish.

(iii) The currents of the system $H[\phi,0]$ under physical (using subscript p) observation are
\begin{eqnarray}
\vec J_{1p}(r)&=& \vec J(r), \label{pc1}\\
\vec J_{2p}(r)&=& \vec J^Q(r)+\phi(r)\circ\vec J(r). \label{pc2} 
\end{eqnarray}
In terms of $\vec J_{1r}(r)$ and $\vec J_{2r}(r)$, and $\Phi_1$ and $\Phi_2$, they read
\begin{eqnarray}
\vec J_{1p}(r)&=& \vec J_{1r}(r)+\vec J_1(r)\circ\Phi_2(r), \nonumber\\
\vec J_{2p}(r)&=& \vec J_{2r}(r)+\vec J_1(r)\circ\Phi_1(r)+2\vec J_2(r)\circ\Phi_2(r). \nonumber
\end{eqnarray}
The observed current densities are their averages in the system $H[\phi,0]$. 

By the linear response theory, we have
\begin{eqnarray}
\langle \vec J_{1p}(r)\rangle &=& \langle\vec J_1(r)\circ\Phi_2(r)\rangle_0 -\hat K^{11}\cdot\nabla\Phi_1 - \hat K^{12}\cdot\nabla\Phi_2,  \nonumber\\
 \label {J1} \\
\langle \vec J_{2p}(r)\rangle &=& \langle\vec J_1(r)\circ\Phi_1(r)\rangle_0+2\langle\vec J_2(r)\circ\Phi_2(r)\rangle_0 \nonumber\\
&&-\hat K^{21}\cdot\nabla\Phi_1 - \hat K^{22}\cdot\nabla\Phi_2,   \label {J2}
\end{eqnarray}
where $\langle \cdots\rangle_0$ is the average under $H[\Phi,\Psi]$, $\hat K^{ij}$'s are $2\times 2$ constant tensors [with respect to the directions of the coordinates hereafter denoted by subscripts $(x,y)$ or $(\mu,\nu)$, the superscripts 1 corresponding to particle current and 2 to heat current] determined by the Kubo formula
\begin{equation}
\hat K^{ij} = -\lim_{\Omega\to 0}{\rm Im}\hat \chi^{ij}(\Omega+i0)/\Omega \label{kij}
\end{equation}
with $\hat\chi^{ij}(\Omega+i0)$ as the retarded response function. In the bosonic Matsubara frequency $\Omega_m$, $\hat \chi^{ij}$ reads
\begin{eqnarray}
\hat\chi^{ij}(i\Omega_m) &=& -\frac{1}{V}\int_{0}^{\beta}d\tau e^{i\Omega_m\tau}\langle T_{\tau}\vec J_i(\tau)\vec J_j(0)\rangle_0 \label{chi}
\end{eqnarray}
where $\vec J_i = \int d\vec r\vec J_i(r)$. Now that $\langle \vec J_{ip}(r)\rangle$ given by Eqs. (\ref{J1}) and (\ref{J2}) are already explicitly linear in the perturbations $\Phi_1$ and $\Phi_2$, in calculating the averages for the constants in Eqs. (\ref{J1}) and (\ref{J2}), the equilibrium point $[\Phi,\Psi]$ can be shifted to $[0,0]$. To the first order of $\nabla\Phi_i$, the term -$\langle\vec J_i(r)\circ\Phi_j(r)\rangle_0$'s in Eqs. (\ref{J1}) and (\ref{J2}) are calculated as 
\begin{eqnarray}
\langle\vec J_i(r)\circ\Phi^j(r)\rangle_0 = \langle \vec J_i(r)\circ(\vec r\cdot\nabla\Phi_j)\rangle_0. \nonumber
\end{eqnarray}
For the diagonal elements $\langle J_{ix}(r)\circ x\rangle_0 \equiv M^i_{xx}$, we have
\begin{eqnarray}
M_{xx} &=& \langle J_{x}(r)\circ x\rangle_0 =i\langle\psi^{\dagger}(r)[\xi(r),x^2]\psi(r)\rangle_0/2 =0,  \nonumber\\
M^Q_{xx} &=&\langle J^Q_{x}(r)\circ x\rangle_0= i\langle\psi^{\dagger}(r)[\xi^2(r),x^2]\psi(r)\rangle_0/4 =0, \nonumber
\end{eqnarray}
which can be shown by expanding $\psi(r)$ in terms of the eigen states of $\xi(r)$. For the off-diagonal elements, $\langle J_{ix}(r)\circ y\rangle_0 \equiv M^i_{xy}$, we get 
\begin{eqnarray}
M^i_{xy} &=& \langle J_{ix}(r)y\rangle_0 ,  \nonumber\\
 &=& \langle [J_{ix}(r)y-J_{iy}(r)x]\rangle_0/2 ,  \nonumber\\
 &=& -M^i_{yx} ,  \label{mxy}
\end{eqnarray}
because the system is invariant under rotation around the $z$-axis perpendicular to the plane. The element $M^i_{xy}$ is the magnetization of the electrons. Substituting the results into Eqs. (\ref{J1}) and (\ref{J2}), we get 
\begin{eqnarray}
\langle \vec J_{1p}(r)\rangle &=& -\hat N^{11}\cdot\nabla\Phi_1 - \hat N^{12}\cdot\nabla\Phi_2, 
 \label {J1a} \\
\langle \vec J_{2p}(r)\rangle &=& -\hat N^{21}\cdot\nabla\Phi_1 - \hat N^{22}\cdot\nabla\Phi_2,   \label {J2a}
\end{eqnarray}
with 
\begin{eqnarray}
\hat N^{11} &=& \hat K^{11} \label {n11} \\
\hat N^{12} &=& \hat K^{12}-\hat M  \label {n12} \\
\hat N^{21} &=& \hat K^{21}-\hat M  \label {n21} \\
\hat N^{22} &=& \hat K^{22}-2\hat M^Q.  \label {n22}
\end{eqnarray}
These forms have been obtained in Ref. \onlinecite{Cooper} with a phenomenological approach using the Einstein argument and the consideration classifying the transport components in each kind of the local currents as the observable currents. 

Though we consider the Dirac fermions here, the derivation above is valid for general electron systems.
 
For a non-interacting system such as the one considered here, the average $\langle\cdots\rangle_0$ under $H[0,0]$ reduces, in principle, to the independent single-particle problem. Using the bases of single particle states $\{|n\rangle\}$ for a given impurity configuration, one can obtain formally the expressions for $\hat N^{11}$ and $\hat N^{12}=\hat N^{21}$ as
\begin{eqnarray}
\hat N^{11} &=& -\int\frac{d\omega}{2\pi}f'(\omega)\hat C(\omega), \label{n1}\\
\hat N^{12} &=& \hat N^{21} = -\int\frac{d\omega}{2\pi}f'(\omega)\omega\hat C(\omega), \label{n2} 
\end{eqnarray}
with
\begin{equation}
\hat C(\omega) = \frac{2\pi^2}{V}\langle\sum_{n\ne m}{\rm Re}\vec j_{nm}\vec j_{mn}\delta(\omega-\xi_n)\delta(\omega-\xi_m)\rangle_i   \label{C}
\end{equation}
where $f'(\omega)=df(\omega)/d\omega$ with $f(\omega)$ the Fermi distribution function, $\vec j_{nm}=\langle n|\vec j|m\rangle$, $\xi_n$ is the eigenvalue of $\xi(r)$ of the state $|n\rangle$ and $\langle\cdots\rangle_i$ in Eq. (\ref{C}) means the average over the impurity configurations. These forms have been obtained in Ref. \onlinecite{Jonson}. For the readers' convenience, we give a derivation in Appendix. The compact form given by Eqs. (\ref{n2}) is so obtained because of large cancellations between $\hat K^{12} = \hat K^{21}$ and $\hat M$. The tensors $\hat N^{11}$ and $\hat N^{12}=\hat N^{21}$ are thus related via the function $\hat C(\omega)$. 

However, Eq. (\ref{C}) is not a convenient formula to start with. To proceed, one needs to derive $\hat C(\omega)$ in terms of the Green's function. From Eq. (\ref{chi}), carrying out the $\tau$-integral, we have
\begin{eqnarray}
\chi^{11}_{\mu\nu}(i\Omega_m) &=& \frac{1}{\beta V}\sum_n\int d\vec r\int d\vec r' {\rm Tr}\langle G(r,r',i\omega_n)J_{1\mu} \nonumber\\
&& \times G(r',r,i\omega_n+i\Omega_m)J_{1\nu}\rangle_i  \nonumber\\
&\equiv& \frac{1}{\beta}\sum_n P_{\mu\nu}(\omega_n,\omega_n+\Omega_m) \label{phi}
\end{eqnarray}
where $G(r,r',i\omega_n)$ is the Green's function for a given impurity distribution, $\omega_n$ is the fermionic Matsubara frequency, $\langle\cdots\rangle_i$ again the average over the impurity configurations, and the function $P_{\mu\nu}(\omega_1,\omega_2)$ is so defined by the equation. Taking the analytical continuation $\Omega_m \to \Omega + i0$, we have for $\hat N^{11}$, 
\begin{eqnarray}
\hat N^{11} &=& -\int\frac{d\omega}{2\pi}f'(\omega){\rm Re}[\hat P(\omega^-,\omega^+)-\hat P(\omega^-,\omega^-)] \nonumber\\
&&+\int\frac{d\omega}{2\pi}f(\omega){\rm Re}\frac{\partial}{\partial\omega'}[\hat P(\omega',\omega^+)-\hat P(\omega^+,\omega')]_{\omega'=\omega^+}\nonumber\\
&\equiv& \int\frac{d\omega}{2\pi}f'(\omega){\rm Re}[\hat P(\omega^-,\omega^-)-\hat P(\omega^-,\omega^+)-\hat Y(\omega^+)] \nonumber
\end{eqnarray} 
where $\omega^{\pm} = \omega \pm i0$, and 
\begin{eqnarray}
\hat Y(\omega) &=& \int_{-\infty}^{\omega}dz\frac{\partial}{\partial z'}[\hat P(z',z)-\hat P(z,z')]_{z'=z}.\nonumber
\end{eqnarray} 
Therefore we get
\begin{eqnarray}
\hat C(\omega) &=& {\rm Re}[\hat P(\omega^-,\omega^+)-\hat P(\omega^-,\omega^-)+\hat Y(\omega^+)]. \nonumber
\end{eqnarray} 

\subsection{Hall and Nernst conductivities of Dirac fermions}

The Nernst effect describes the response of the transverse current to a temperature gradient $\nabla T$ in the presence of a perpendicular magnetic field $B$ but $\vec E =0$. It is reflected by the Nernst conductivity $N^{12}_{xy}$. Here, we study it in the limit of $B \to 0$.

In the limit of $B \to 0$, the elements $N^{11}_{xx}$ and $N^{12}_{xx}$ are independent of the magnetic field. They are related to the electric conductivity $\sigma = e^2N^{11}_{xx}$ and thermoelectric power $S = -N^{12}_{xx}/eTN^{11}_{xx}$ which we have given in previous works.\cite{Yan1,Yan2} Though the element $N^{11}_{xy}$ was calculated previously for studying the Hall coefficient, the function $C_{xy}(\omega)$ was not given explicitly.\cite{Yan1} To calculate $N^{12}_{xy}$, we here need to find out the explicit expression for $C_{xy}(\omega)$. 

\begin{figure} 
\centerline{\epsfig{file=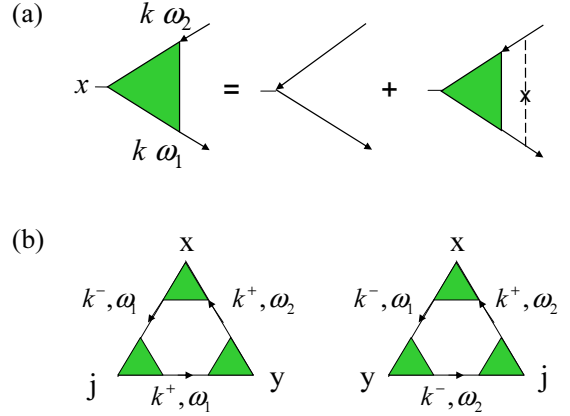,width=8. cm}}
\caption{(Color online) (a) Self-consistent Born approximation for the Current vertex correction. (b) Diagrams for calculating the function $P_{xy}(\omega_1,\omega_2)$. The vertex $j$ is associated with the vector potential $A_j$. $k^{\pm} = k \pm q/2$. } 
\end{figure} 

As in the calculation of the Hall conductivity in the limit of $B \to 0$,\cite{Yan1} by introducing the vector potential via $\vec A(r) = \vec A(q)\exp(i\vec q\cdot\vec r)$ with $\vec B = i\vec q\times\vec A(q)$ and taking the limit $q \to 0$, one obtains $\Phi^{12}_{xy}(i\Omega_m)$ in terms of the average of the multiplication of three current operators.\cite{Fukuyama,Nakamura} As shown in Fig. 1(b), $P_{xy}(\omega_1,\omega_2)$ is obtained as
\begin{eqnarray}
P_{xy}(\omega_1,\omega_2) &=& \lim_{q\to 0}\frac{2e}{V}\sum_{k}{\rm Tr}\{\Gamma_x(k^-,k^+,\omega_1,\omega_2)\nonumber\\
&&\times[G(k^+,\omega_2)\Gamma_y(k^+,\omega_2,\omega_1)V(k^+,k^-,\omega_1)  \nonumber\\
&&+V(k^+,k^-,\omega_2)\Gamma_y(k^-,\omega_2,\omega_1)G(k^-,\omega_1)] \}  \nonumber
\end{eqnarray}
where the factor 2 stems from the spin degeneracy, $k^{\pm} = k\pm q/2$, $V(k^+,k^-,\omega)= G(k^+,\omega)\vec\Gamma(k^+,k^-,\omega,\omega)\cdot\vec AG(k^-,\omega)$, and  $\Gamma_x(k,\omega_1,\omega_2)\equiv \Gamma_x(k,k,\omega_1,\omega_2)$ as given by Eq. (\ref{vt}). The vertex $\Gamma_{\mu}(k^-,k^+,\omega_1,\omega_2)$ satisfies the $4\times 4$ matrix equation
\begin{eqnarray}
\Gamma_{\mu}(k^-,k^+,\omega_1,\omega_2) &=& \tau_3\sigma_{\mu}+\frac{1}{V}\sum_{k_1}n_iv^2_0(k-k_1) G(k^-_1,\omega_1)\nonumber\\
&&\times\Gamma_{\mu}(k^-_1,k^+_1,\omega_1,\omega_2)G(k^+_1,\omega_2). \label{vt2} 
\end{eqnarray} 
To find out the limit of $q\to 0$, we need to expand the right hand side of Eq. (\ref{phi}) to the first order in $q$ and then use $\vec B = i\vec q\times\vec A$. The manipulation is tedious but elementary. We only outline the key points in the derivation below.

(i) The expansions of the Green function $G(k^{\pm},\omega)$ and the vertex functions $\Gamma_{\mu}(k^{\pm},\omega_1,\omega_2)$ can be easily obtained by definition. The most involved expansion is for the vertex function $\Gamma_x(k^-,k^+,\omega_1,\omega_2)$. By expanding both sides of Eq. (\ref{vt2}) to the first order in $q$, one gets  $\Gamma_{\mu}(k^-,k^+,\omega_1,\omega_2)$ = $\Gamma_{\mu}(k,\omega_1,\omega_2)+ \vec\gamma_{\mu}(k,\omega_1,\omega_2)\cdot\vec q/2$ with $\vec\gamma_{\mu}(k,\omega_1,\omega_2)$ determined by
\begin{eqnarray}
\vec\gamma_{\mu}(k,\omega_1,\omega_2)&=&\frac{1}{V}\sum_{k'}n_iv^2_0(k-k') G(k',\omega_1)\vec\gamma_{\mu}(k',\omega_1,\omega_2) \nonumber\\
&&\times G(k',\omega_2)-\frac{1}{V}\sum_{k'}n_iv^2_0(k-k') \nonumber\\
&& \times[\nabla G(k',\omega_1)\Gamma_{\mu}(k',\omega_1,\omega_2)G(k',\omega_2) \nonumber\\
&&-G(k',\omega_1)\Gamma_{\mu}(k',\omega_1,\omega_2)\nabla G(k',\omega_2)] \nonumber\\ 
\label{dw}
\end{eqnarray} 
where $\nabla$ means the gradient with respect to $k'$. 

(ii) From the identity
\begin{eqnarray}
&&\frac{1}{V}\sum_{kk'}n_i[(\nabla_k+\nabla_{k'}) v^2_0(k-k')] {\rm Tr}[G(k,\omega_1)\vec\gamma_{\mu}(k,\omega_1,\omega_2)\nonumber\\
&& \times G(k,\omega_2)G(k',\omega_2)\Gamma_{\nu}(k',\omega_2,\omega_1)G(k',\omega_1)] = 0, \label{id}
\end{eqnarray} 
performing the integral by part in the left hand side of Eq. (\ref{id}) and using Eq. (\ref{dw}) and the equation for $\Gamma_{\nu}(k,\omega_1,\omega_2)$, we obtain
\begin{eqnarray}
&&\sum_k{\rm Tr}\{\vec\gamma_{\mu}(k,\omega_1,\omega_2)[\nabla_kG(k,\omega_2)\Gamma_{\nu}(k,\omega_2,\omega_1)G(k,\omega_1)\nonumber\\
&&+ G(k,\omega_2)\Gamma_{\nu}(k,\omega_2,\omega_1)\nabla_kG(k,\omega_1)]\}\nonumber\\
&=& -\sum_{k}{\rm Tr}\{[\nabla G(k,\omega_1)\Gamma_{\mu}(k,\omega_1,\omega_2)G(k,\omega_2)  \nonumber\\
&&- G(k,\omega_1)\Gamma_{\mu}(k,\omega_1,\omega_2)\nabla G(k,\omega_2)]\nabla_k \Gamma_{\nu}(k,\omega_2,\omega_1)\}. \nonumber
\end{eqnarray} 

(iii) For $V(k^+,k^-,\omega)$, using Eq. (\ref{vt2}), we get the expansion
\begin{eqnarray}
&&G(k^+,\omega)\Gamma_{\alpha}(k^+,k^-,\omega,\omega)G(k^-,\omega) \nonumber\\
&=& \frac{\partial}{\partial k_{\alpha}} G(k,\omega)+i\sigma_z\hat\alpha\cdot[b(k,\omega)\hat k\hat\phi-a(k,\omega)\hat\phi\hat k]\cdot\vec q. \nonumber\\
\label{we1}
\end{eqnarray} 
where $\phi$ is the angle of $\vec k$, $\hat \phi$ is the unit vector in $\phi$ direction, and the first term in the right hand of Eq. (\ref{we1}) comes from the Ward identity $G(k,\omega)\Gamma_{\alpha}(k,\omega,\omega)G(k,\omega) = \partial G(k,\omega)/\partial k_{\alpha}$. The coefficients $a(k,\omega)$ and $b(k,\omega)$ are determined by solving Eq. (\ref{dw}). Since the final result depends on their combination $a(k,\omega)+b(k,\omega)\equiv c(k,\omega)$, the function $c(k,\omega) \equiv z(k,\omega)+[g^2_0(k,\omega)-g^2_c(k,\omega)]X(k,\omega)$ is determined by following equations: 
\begin{eqnarray}
z(k,\omega) &=& [g'_0(k,\omega)g_c(k,\omega)-g_0(k,\omega)g'_c(k,\omega)]\nonumber\\
&&\times[y_0(k,\omega,\omega)-y_3(k,\omega,\omega)] \nonumber\\
&&-g_c(k,\omega)\{g_0(k,\omega)[y_0(k,\omega,\omega)+y_3(k,\omega,\omega)]\nonumber\\
&&+2g_c(k,\omega)y_1(k,\omega,\omega)\}/k ,\nonumber\\
X(k,\omega) &=& \frac{1}{V}\sum_{k'}n_iv^2_0(k-k')\{z(k',\omega)\nonumber\\
&&+g_+(k',\omega)g_-(k',\omega)X(k',\omega)\}, \nonumber
\end{eqnarray} 
with $g'(k,\omega)=\partial g(k,\omega)/\partial k$. 

Using the results given above, one gets a final expression $P_{xy}(\omega_1,\omega_2) = [Z(\omega_1,\omega_2)-Z(\omega_2,\omega_1)]/2i $ with
\begin{eqnarray}
Z(\omega_1,\omega_2) &=& \frac{2Be}{V}\sum_k{\rm Tr}[G_1
\Gamma_{x,12}(\frac{\partial G_2}{\partial k_x}\frac{\partial\Gamma_{y,21}}{\partial k_y} \nonumber\\
&&-\frac{\partial G_2}{\partial k_y}\frac{\partial\Gamma_{y,21}}{\partial k_x}) 
-ic(k,\omega_1)\Gamma_{x,12}G_2\Gamma_{y,21}\sigma_z] \nonumber
\end{eqnarray} 
where $\Gamma_{x,ij} = \Gamma_x(k,\omega_i,\omega_j)$ (and the same meaning for $\Gamma_{y,ij}$), $G_j = G(k,\omega_j)$. Since $P_{xy}(\omega^-,\omega^-) = 0$, we have $C_{xy}(\omega) = {\rm Im}Z(\omega^-,\omega^+)+{\rm Re} Y_{xy}(\omega^+)$, and
\begin{eqnarray}
{\rm Re} Y_{xy}(\omega) &=& \int_{-\infty}^{\omega}dz {\rm Im}R(z) \nonumber
\end{eqnarray} 
with $R(\omega)=\frac{\partial}{\partial\omega'}[Z(\omega',\omega)- Z(\omega,\omega')]_{\omega'=\omega}$. 

Knowing the function $C_{xy}(\omega)$, we can calculate $N^{11}_{xy}$ and the Nernst conductivity $N^{12}_{xy}$ according to Eqs. (\ref{n1}) and (\ref{n2}), respectively. Since we are interested in the low temperature cases, we here give their expressions in the limit of $T \to 0$. They are
\begin{eqnarray}
N^{11}_{xy} &=& \frac{1}{2\pi}{\rm Im}Z(0^-,0^+) 
-\int_0^{\infty}\frac{dz}{2\pi}{\rm Re}R(iz),\label{Lxy}\\
N^{12}_{xy} &=& \frac{\pi T^2}{6}{\rm Im}[\frac{\partial}{\partial\omega}Z(\omega^-,\omega^+)|_{\omega =0}
+R(0^+)],\label{Nxy}
\end{eqnarray} 
In obtaining $N^{12}_{xy}$, the use of the expanding $C_{xy}(\omega) \approx C_{xy}(0)+\omega C'_{xy}(0)$ has been made. In addition, the $z$-integral in Re$Y_{xy}(0)$ reduces to the path-integral around the negative axis $(-\infty, 0)$. Because $R(z)$ is an analytical function (by definition) in the upper and lower z-plane, respectively, the integral path has been deformed to the imaginary axis, giving rise to the last term in Eq. (\ref{Lxy}). The integral along the imaginary axis is simple to handle for the numerical calculation since there is no singularity in the Green's function. If this term is neglected, the expression for the Hall conductivity $\sigma_{xy} = e^2N^{11}_{xy}$ will reduce to the form as in the previous work.\cite{Yan1} The fact that the contribution from this term is negligible will be checked later. In analogous to the thermoelectric power $S = E_x/(\nabla T)_x$, we define $S_{xy} = E_x/(\nabla T)_y$ which describes the production of the transverse electric field due to a temperature gradient in the absence of current flow. By setting $\langle J_{1p}(r)\rangle = 0$ in Eq. (\ref{J1a}), $S_{xy}$ is obtained as
\begin{eqnarray}
S_{xy} = -(S\sigma_{xy}+eN^{12}_{xy}/T)/\sigma.\label{Sxy}
\end{eqnarray} 
Clearly, $S_{xy}$ includes two parts. The quantity $S\sigma_{xy}$ describe the process for the response of the transverse electric field to $\nabla T$ through the way: Because there exist a longitudinal electric field (with the magnitude proportional to the thermoelectric power $S$) due to the temperature gradient, the current could flow transversely by the Hall process. In different to this, $eN^{12}_{xy}/T$ implies an additional transverse field due to a transverse current response directly to $\nabla T$.

\section{Results}
 
The functions $g_{0,c}(k,\omega)$ and $y_j(k,\omega_1,\omega_2)$ and their derivatives with respect to $k$ and $\omega$ are involved in the calculation. In a recent work,\cite{Yan2} we have described how to numerically solve the corresponding integral equations to determine these functions. In our numerical calculation, we take the impurity density as $n_i = 1.15\times 10^{-3}a^{-2}$ (with $a$ as the lattice constant of graphene) the same as in our previous works.\cite{Yan1,Yan2,Yan} 
 
\begin{figure}
\centerline{\epsfig{file=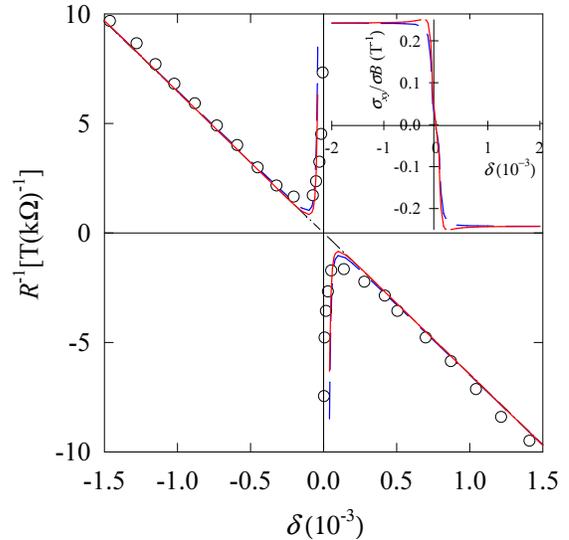,width=8. cm}} 
\caption{(Color online) The inverse Hall coefficient $R^{-1}$ (in unit of $10^{-3}$Tesla/Ohm) as a function of $\delta$. The solid (red) and dashed (blue) lines correspond respectively to the calculations with and without the contribution from the last term in Eq. (\ref{Lxy}) [Inset, the corresponding results for $\sigma_{xy}$ normalized by $\sigma B$ with $B$ is in unit of Tesla (T)]. Dot-dashed line: classical result. Symbols: experimental data (Ref. \onlinecite{Geim}). } 
\end{figure} 

The numerical results for the Hall conductivity $\sigma_{xy}$ and the inverse Hall coefficient $R^{-1} = B\sigma^2/\sigma_{xy}$ as functions of the carrier concentration $\delta$ (doped carrier per carbon atom) are shown in Fig. 2. The calculations with and without the last term in Eq. (\ref{Lxy}) for both results of $R^{-1}$ and  $\sigma_{xy}$ (normalized by $\sigma B$ in inset of Fig. 2) are almost indistinguishable. This fact means that the contribution from the last term in Eq. (\ref{Lxy}) is negligible small. The experimental data \cite{Geim} (symbols) and the classical prediction $R^{-1} = -nec$ are also plotted in Fig. 2 for comparison. As we have stated previously, the divergence of $R^{-1}$ at $\delta =0$ stems from the vanishing of $\sigma_{xy}$ while the conductivity $\sigma$ remains finite. 

At low $T$, the Nernst conductivity $N^{12}_{xy}$ is proportional to $T^2$. In Fig. 3, we depict $eN^{12}_{xy}/T^2B$ as a function of the electron doping concentration $\delta$. Because of the electron-hole symmetry, $N^{12}_{xy}$ is even for $\delta \to -\delta$. $N^{12}_{xy}$ comes mainly from the first term in the square bracket in Eq. (\ref{Nxy}). This is similar to the case as in $N^{11}_{xy}$. The derivation $\partial Z(\omega^-,\omega^+)/\partial\omega|_{\omega=0}$ is very delicate. Here, it cannot be considered simply as $\propto \partial \sigma_{xy}(\mu)/\partial\mu$ because the scattering potential here depends strongly on the electron doping concentration. For comparison, we also depict the result for $S\sigma_{xy}/TB$ in Fig. 3. In the regime of $\delta$ studied here, the magnitudes of both quantities are overall the same. But at low $\delta$, $eN^{12}_{xy}/T^2B$ is biger than $S\sigma_{xy}/TB$. In the absence of the electric field, the transverse current density is solely determined by $J_x = -N^{12}_{xy}(\nabla T)_y/T$.  

\begin{figure} 
\centerline{\epsfig{file=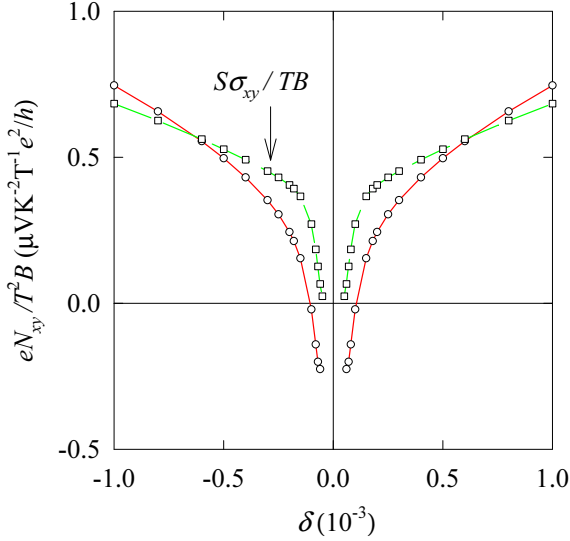,width=8. cm}}
\caption{(Color online) The coefficient $eN^{12}_{xy}/T^2B$ (red line with circles)(in unit of $\mu VK^{-2}T^{-1}e^2/h$ with $h$ the Planck constant) as a function of $\delta$. The dashed line (green with squares) is $S\sigma_{xy}/TB$ for comparison.} 
\end{figure} 

In Fig. 4, we show the transverse thermoelectric power $S_{xy}$ divided by $TB$ as functions of $\delta$ at the average impurity densities $n_i = 1.15\times 10^{-3}a^{-2}$ and $n_i = 1.6\times 10^{-3}a^{-2}$. $S_{xy}$ is a superposition of $S\sigma_{xy}$ and $eN^{12}_{xy}/T$. As a result, $S_{xy}$ is linear in $T$ and $B$. At large carrier doping, both of $S\sigma_{xy}$ and $eN^{12}_{xy}/T$ have about the same contribution to $S_{xy}$. While at low doping, $eN^{12}_{xy}/T$ is predominant. $S_{xy}$ changes sign at low doping because $eN^{12}_{xy}/T$ does. The factor $S\sigma_{xy}+eN^{12}_{xy}/T$ of $S_{xy}$ decreases quickly as $\delta$ decreasing at low $\delta$ regime. Its slop normalized by a negative constant ($\sim -\sigma/\delta$, $\sigma \propto \delta$ at large $\delta$ and is flat at very low doping) is almost the behavior of $S_{xy}$. The dip in $S_{xy}$ corresponding to the maximum of the slop. So far there exist no measurements of the Nernst effect of Dirac fermions in a weak magnetic field. But for the purpose to make a qualitative comparison, the experimental results for the transverse thermoelectric power of Ref. \onlinecite{Wei} with the magnetic fields $B$ = 1 Tesla (T) (squares) and 3 T (diamonds) are plotted.  Here the ratio of the first non-zero Landau level/the Fermi energy is about 0.35 for a doping level at $1.0\times 10^{-4}$ and $B$ = 1 T. Though the strengths of these magnetic fields could not be regarded as weak, the data indicate a tendency toward to the theoretical prediction as the magnetic field is decreasing.   

The present calculation is based on the assumption that the impurities are randomly distributed. Actually, at low carrier doping, graphene is an inhomogeneous system due to the impurity correlations in the substrate as observed by experiment.\cite{Martin,Zhang2,Berezovsky,Deshpande} It seems there exist the electron and hole puddles. Nonetheless, in each puddle, the average number of the carrier is less than unity. Moreover, the mean free path of the electrons is much longer than the length scale ($\sim$ tens nanometers) of the puddles. Within a mean free path, a carrier can transfer through many such puddles. The puddles can be thus regarded as the microscopic wrinkles. In addition, at low carrier concentration, there exists significant quantum coherence between the upper- and lower-band states,\cite{Trushin} resulting in the minimum electric conductivity,\cite{Shon,Peres,Ostrovsky,Yan} the unconventional behaviors of the inverse Hall conductivity\cite{Geim,Yan1} and the thermoelectric power.\cite{Zuev,Wei,Ong,Yan2} Therefore, the carrier must be treated quantum mechanically and the present approach seems plausible. 
 
\begin{figure} 
\centerline{\epsfig{file=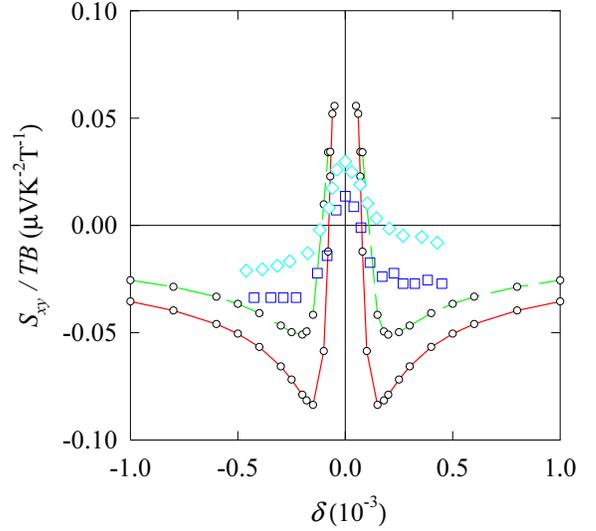,width=8. cm}}
\caption{(Color online) The coefficient $S_{xy}/TB$ of transverse thermoelectric power (in unit of $\mu VK^{-2}T^{-1}$) as a function of $\delta$. The solid (red with small symbols) and dashed (green) lines are the calculations at impurity densities $n_i = 1.15\times 10^{-3}a^{-2}$ and $1.6\times 10^{-3}a^{-2}$, respectively. The squares (blue at $B = 1$ Tesla) and diamonds (cyan at $B = 3$ Tesla) are the experimental results of Ref. \onlinecite{Wei}. } 
\end{figure} 

Finally, we compare our calculation with the semi-classical Boltzmann theory. By the Boltzmann theory within the relaxation time-$\tau$ approximation, the function $g_k$ describing the difference between the disturbed distribution function and the Fermi distribution function $f$ is determined by\cite{Ziman}
\begin{eqnarray}
g_k &=& \tau\frac{\partial f}{\partial\xi_k}\vec v\cdot(\frac{\xi_k}{T}\nabla T+e\vec E)
+\tau e(\vec v\times\vec B)\cdot \nabla_kg_k  \nonumber\\
\label{Bltz}
\end{eqnarray}
using units of $c = \hbar = 1$ again. Here $\xi_k = vk-\mu$ for electrons in graphene. From Eq. (\ref{Bltz}), one obtains
\begin{eqnarray}
\sigma &=& \tau v e^2k_F/\pi   \nonumber\\
\sigma_{xy} &=& B\tau^2 v^2e^3 /\pi   \nonumber\\
N^{12}_{xx} &=& T^2\tau \pi /3.  \nonumber
\end{eqnarray}
The inverse Hall coefficient is $B\sigma^2/\sigma_{xy}= ne$. The Matt relation\cite{Mott} is given by $N^{12}_{xx} = T^2\pi^2\sigma'/3e^2$ with $\sigma' = \tau e^2/\pi$. However, using the Mott relation, one obtains zero Nernst conductivity $N^{12}_{xy} = 0$ because $\sigma_{xy}$ is constant independent of the chemical potential $\mu$. This is different from the present result that the Nernst conductivity is in the order of $S\sigma_{xy}T/e$ as shown in Fig. 3. 

\section{Summary}

In summary, we have derived the formula for the transport coefficients for the Dirac fermions in graphene in the presence of the temperature gradient, the electric field and the magnetic field. The derivation is valid for general electron systems. It is different from the usual perturbation process with the external dynamic potentials applied (in that case the original equilibrium state is unchanged) since the perturbation due to turning on the temperature gradient shifts the equilibrium state. The physical observed system is in the perturbed state with respect to this equilibrium state.

On the basis of self-consistent Born approximation, we have studied the Nernst effect of the Dirac fermions under the charged impurity scatterings and weak magnetic field in graphene. The transverse thermoelectric power is closely related with the Hall conductivity and the longitudinal thermoelectric power for which the theory has been shown to be in good agreement with the experiment. The Nernst conductivity is dealt with the similar approach as for the Hall conductivity. The present calculation is a prediction to the Nernst conductivity of the Dirac fermions in graphene under a weak magnetic field.

\vskip 5mm
\centerline {\bf Acknowlegements}
\vskip 3mm

This work was supported by a grant from the Robert A. Welch Foundation under Grant No. E-1146, the TCSUH, the National Basic Research 973 Program of China under Grant No. 2005CB623602, NSFC under Grants No. 10774171 and No. 10834011, and a financial support from the Chinese Academy of Sciences for advanced researches.

\vskip 5mm
\centerline {\bf APPENDIX}
\vskip 3mm

We here give a derivation of Eqs. (\ref{n1}) and (\ref{n2}). Consider the response function $\chi_{AB}(\tau)$ defined by
\begin{eqnarray}
\chi^{AB}(\tau) = -\frac{1}{V}\langle T_{\tau}A(\tau)B(0)\rangle_0 \label{ab}
\end{eqnarray}
where $A$ and $B$ (of Hermitians) can be either the current or the heat current operator. Using the basis of the single particle states $\{|n\rangle\}$, we have 
\begin{eqnarray}
A(\tau) &= &\int d\vec r e^{H[0,0]\tau}\psi^{\dagger}(r)a(r)\psi(r) e^{-H[0,0]\tau} \nonumber\\
&= &\sum_{nm}e^{(\xi_n-\xi_m)\tau}a_{nm}c^{\dagger}_nc_m \label{a1}
\end{eqnarray}
where $a_{nm} = \langle n|a|m\rangle$ and $c^{\dagger}_n$ ($c_n$) creates (annhilates) a particle in state $|n\rangle$. The function $\chi_{AB}(\tau)$ is written as 
\begin{eqnarray}
\chi^{AB}(\tau) &= &-\frac{1}{V}\langle\sum_{n}f_na_{nn}b_{nn} \nonumber\\
&&+\sum_{n\ne m}f_n(1-f_m)e^{(\xi_n-\xi_m)\tau}a_{nm}b_{mn} \rangle_i\nonumber
\end{eqnarray}
where $f_n = \langle c^{\dagger}_nc_n\rangle$ is the Fermi distribution function. Taking the Fourier transform, we get
\begin{eqnarray}
\chi^{AB}(i\Omega_{\ell}) &= &-\frac{1}{V}\langle\sum_nf_na_{nn}b_{nn}\beta\delta_{\ell,0} \nonumber\\
&&-\sum_{n\ne m}\frac{f_n-f_m}{i\Omega_{\ell}+\xi_n-\xi_m}a_{nm}b_{mn} \rangle_i.\label{a3}
\end{eqnarray}
where $\beta = 1/T$. Taking the analytical continuation $i\Omega_{\ell} \to \Omega + i0$ [for which the first term in the braces in Eq. (\ref{a3}) can be disregarded because of $\Omega_{\ell} > 0$], we have 
\begin{eqnarray}
{\rm Im}\chi^{AB,r}(\Omega) &= &\frac{1}{V}{\rm Im}\langle\sum_{n\ne m}\frac{f_n-f_m}{\Omega+\xi_n-\xi_m+i0}a_{nm}b_{mn}\rangle_i\nonumber\\ 
&=& \frac{1}{V}\langle\sum_{n\ne m}(f_n-f_m)[\frac{{\rm Im}(a_{nm}b_{mn})P}{\Omega+\xi_n-\xi_m}\nonumber\\
&& -\pi{\rm Re}(a_{nm}b_{mn})\delta(\Omega+\xi_n-\xi_m)]\rangle_i \nonumber\\
&=& \frac{1}{V}\langle\sum_{n\ne m}(f_n-f_m)[\frac{{\rm Im}(a_{nm}b_{mn})\Omega P}{\Omega^2-(\xi_n-\xi_m)^2}\nonumber\\
&& -\pi{\rm Re}(a_{nm}b_{mn})\delta(\Omega+\xi_n-\xi_m)]\rangle_i.\label{a4}
\end{eqnarray}
where $P$ means taking the principle value in the summation, and in the $P$ term the exchange $n \leftrightarrow m$ and the use of Im$(a_{mn}b_{nm})$ = -Im$(a^{\dagger}_{nm}b^{\dagger}_{mn})$ = -Im$(a_{nm}b_{mn})$ have been made in the last equality. Substituting the result of Eq. (\ref{a4}) into Eq. (\ref{kij}), we obtain 
\begin{eqnarray}
K^{AB}&=& \frac{1}{V}\langle\sum_{n\ne m}[{\rm Im}(a_{nm}b_{mn})\frac{(f_n-f_m) P}{(\xi_n-\xi_m)^2}  \nonumber\\
&& -\pi{\rm Re}(a_{nm}b_{mn})f'_n\delta(\xi_n-\xi_m)]\rangle_i  \label{a5}
\end{eqnarray}
where $f'_n = df_n/d\xi_n$. 

Note first, the desired forms given by Eqs. (\ref{n1}) and (\ref{n2}) are obtained from the contribution from the last term in the square brackets in Eq. (\ref{a5}).

Second, for the diagonal elements, the first term in the square brackets in Eq. (\ref{a5}) vanishes. In the present case, $A$ and $B$ are vectors. For the diagonal elements, $a_{x,nm} \propto b_{x,nm}$ and $a_{x,nm}b_{x,mn}$ is real. 

Therefore, in following, we will consider only the contribution from the first term in the square brackets in Eq. (\ref{a5}) for the case of off-diagonal elements. Denote it as
\begin{eqnarray}
R(A,B)&=& \frac{1}{V}\sum_{n\ne m}\frac{(f_n-f_m) P}{(\xi_n-\xi_m)^2}{\rm Im}(a_{nm}b_{mn}) \label{aa}
\end{eqnarray}
dropping the symbol of the average $\langle\cdots\rangle_i$ for briefness. We only need to prove that the off-diagonal element of $R(A,B)$ is cancelled by the corresponding matrix element of $M^i_{xy}$ given by Eq. (\ref{mxy}).

(i) $A = B = \vec J$. Using $\vec j = i[\xi(r),\vec r]$, $R(\vec J,\vec J)$ reads
\begin{eqnarray}
R(J_x,J_y)&=& \frac{1}{V}\sum_{n\ne m}(f_n-f_m){\rm Im}(x_{nm}y_{mn}) \nonumber\\
&=& \frac{1}{V}\sum_{n}f_n{\rm Im}\langle n|[x,y]|n\rangle  \nonumber\\
&=& 0.  \nonumber 
\end{eqnarray}

(ii) $A = \vec J$ and $B = \vec J^Q$ with $\vec J^Q(r) = \{\vec j(r),\xi(r)\}/2$. Using $\vec j^Q_{nm}=\vec j_{nm}(\xi_n+\xi_m)/2=i\vec r_{nm}(\xi_n^2-\xi^2_m)/2$, one gets 
\begin{eqnarray}
R(J_x,J^Q_y) &=& \frac{1}{2V}\sum_{n\ne m}(\xi_n+\xi_m)(f_n-f_m){\rm Im}(x_{nm}y_{mn}).  \nonumber
\end{eqnarray}
On the other hand, we note
\begin{eqnarray}
-M_{xy} &=& -\frac{1}{2V}\sum_{nm}f_n[j_{x,nm}y_{mn}-j_{y,nm}x_{mn}] \nonumber\\
&=& \frac{1}{V}\sum_{nm}f_n(\xi_n-\xi_m){\rm Im}(x_{nm}y_{mn}) \nonumber\\
&=& \frac{1}{2V}\sum_{nm}(f_n+f_m)(\xi_n-\xi_m){\rm Im}(x_{nm}y_{mn}),  \nonumber
\end{eqnarray}
where we have made use of  the exchange $n \leftrightarrow m$ in the last equality. Therefore, we have 
\begin{eqnarray}
R(J_x,J^Q_y)-M_{xy} 
&=& \frac{1}{V}\sum_{nm}(f_n\xi_n-f_m\xi_m){\rm Im}(x_{nm}y_{mn})\nonumber\\
&=& \frac{1}{V}\sum_{n}f_n\xi_n{\rm Im}\langle n|[x,y]|n\rangle \nonumber\\
&=& 0.  \label{a9}
\end{eqnarray}

(iii) The case of $A = \vec J^Q$ and $B = \vec J$ is the same as (ii) and $N^{12}=N^{21}$.

\end{document}